\documentclass[prl,twocolumn,showpacs]{revtex4}
\newcommand{\figurewidth}{1.\columnwidth}

\usepackage{graphics}
\usepackage{amsmath}

\newcommand{\be}{\begin{equation}}
\newcommand{\ee}{\end{equation}}

\begin{document}

\title{Diffusion-limited aggregation in channel geometry}

\author{Ell\'ak Somfai}
\email{e.somfai@warwick.ac.uk}
\affiliation{Department of Physics, University of Warwick,
	Coventry CV4 7AL, United Kingdom}

\author{Robin C. Ball}
\affiliation{Department of Physics, University of Warwick,
	Coventry CV4 7AL, United Kingdom}

\author{Jason P. DeVita}
\affiliation{Michigan Center for Theoretical Physics, Department of Physics,
	University of Michigan, Ann Arbor, Michigan, 48109-1120}

\author{Leonard M. Sander}
\affiliation{Michigan Center for Theoretical Physics, Department of Physics,
	University of Michigan, Ann Arbor, Michigan, 48109-1120}

\date{\today}

\begin{abstract}
We performed extensive numerical simulation of diffusion-limited aggregation
in two dimensional channel geometry.  Contrary to earlier claims, the
measured fractal dimension $D=1.712\pm0.002$ and its leading correction to
scaling are the same as in the radial case.  The average cluster, defined as
the average conformal map, is similar but not identical to Saffman-Taylor
fingers.
\end{abstract}

\pacs{61.43.Hv}

\maketitle


Diffusion-limited aggregation (DLA) has attracted considerable attention
since its introduction by Witten and Sander in 1981 \cite{witten81}.  In
this model an aggregate or cluster grows by capturing diffusing particles
which irreversibly attach to it on first contact.  This is the discrete
model of a wide variety of physical systems in the Laplacian growth class.
This class can be modeled with a field, satisfying the Laplace equation
outside of a growing cluster, where the cluster grows in proportion to the
gradient of the field at the boundary.  A generalization of DLA, known as
the dielectric breakdown model (DBM) \cite{niemeyer84},
allows growth proportional to the
field gradient to the exponent $\eta$.  DLA is regained for $\eta=1$.  
The majority of our current knowledge about DLA is numerical (mostly in
two dimensions and radial geometry), although progress has been made in
the theoretical front as well (see eg.  \cite{ball02prl,ball03pre}).

One of the controversies surrounding DLA in channel geometry has been that the
fractal dimension might differ from that in the radial case. This claim has
been based on small size simulations \cite{meakin86,argoul88,evertsz90} or
small size calculations \cite{kol00,kol01}. In this paper, based on extensive
numerical simulations of off-lattice DLA in channel, we show that the fractal
dimension is asymptotically the \emph{same} in the two geometries.

One of the most important differences between the two geometries is that for
the channel the continuum version of the problem (Laplacian growth) has a
stable solution without tip splitting instability and finger competition.
These stationary translating solutions ---called Saffman-Taylor fingers
\cite{saffman58}--- have been studied in viscous fingering
experiments in Hele--Shaw cells \cite{thome89}.  In this paper we compare them
with average profiles of DLA clusters.


Our first  method for generating DBM in a channel is to use iterated conformal
maps.  The conformal mapping method for the radial case is described in
\cite{hastings98} and \cite{hastings01prl}.  For radial DLA, a map is
created from the unit circle in the $w$ (``mathematical'') plane
to the unit circle with a bump at a randomly
chosen angle in the physical plane; the composition of such maps is 
a map from the unit circle
the DLA cluster.  To produce  DBM clusters, instead of choosing
the angles at random, we use a
monte-carlo method to select bump sites with the correct distribution
\cite{hastings01prl}.  

To adapt this method to a channel, we modify the map of
Stepanov and Levitov \cite{stepanov01} for both periodic and reflective
boundary conditions by requiring the map to be symmetric about the real
axis.  The map is given by

\be 
f_{\Lambda ,\theta}(w) = ln \bigl[g^{-1}(\tilde{f}_{\Lambda} (g(w)))\bigr] , 
\ee 

\noindent where $g(w) = \frac{w-1}{w+1}$ is a map from the unit circle to
the imaginary axis, and from the exterior of the unit circle to the
positive-real half-plane.  The function

\be\label{ftilde} 
{\tilde f}_{\Lambda}(w) = \frac{w\, +\, \gamma \sqrt{(w-xi)^2+\Lambda^2}\;  
+\, \gamma \sqrt{(w+xi)^2+\Lambda^2} }{ 1\, +\, \gamma 
\sqrt{(1-xi)^2+\Lambda^2}\; +\, \gamma \sqrt{(1+xi)^2+\Lambda^2}} 
\ee

\noindent adds two bumps at symmetric points $x^\prime i$ and $-x^\prime
i$.  The denominator in
Eq.~(\ref{ftilde}) forces ${\tilde f}_{\Lambda}(w)$ to map $w=1$ to
$1$, so that $f_{\Lambda ,\theta}(w)$ maps infinity to infinity.  The
parameter $\Lambda$ controls the size of the bump, and $\gamma$ controls 
the aspect ratio of the bump.
Since $g(w)$ maps $-1$ to infinity, we choose  $\theta$ 
at random between $0$ and $\pi$; and for $\theta > \frac{\pi}{2}$,
$f_{\Lambda ,\theta}(w)$ becomes

\begin{equation}
f_{\Lambda ,\theta}(w) = -\overline{ g^{-1} (\tilde{f}_{\Lambda} 
(g(-\bar{w}))) }
\end{equation}
where the bar denotes complex conjugation.  

The actual bump positions are not at $\pm xi$, but are off by a small 
factor determined by $x$ and $\Lambda$.  The bump size is also dependent 
on $x$ and $\Lambda$.  In order to get bumps at angle $\theta$, ${\tilde 
f}_{\Lambda}$ must place bumps at $g(e^{\pm i\theta}) = 
\frac{sin \theta}{1+cos \theta}i$. We do this by an approximation method.  
To keep all the particles in the cluster the same 
size, the bump size on the unit circle is varied according to the first 
derivative of the composite conformal map in the original version of the
conformal map technique \cite{hastings98}  
This assumes that the higher 
order derivatives are negligible, which is not true deep inside a fjord.  
Thus particles added in a fjord can end up being very large, and sometimes 
can partially fill the channel.  To combat this effect, we measure the 
bump area at each step, and iteratively correct the size parameter 
$\Lambda$ if the area is outside of a preset tolerance (10\% for the 
results in this paper); compare \cite{stepanov01}.

While conformal mapping allows one to grow DBM for any $\eta$, and directly 
produces a conformal map for the cluster boundary, it is computationally 
intensive.  A more efficient numerical algorithm for generating 
off-lattice DLA (that is, $\eta=1$) is a simple adaptation
of hierarchical maps \cite{ball85} to channel geometry.  This method enables
close to linear dependence of computing resources on cluster size. We used in
total $1.7\times 10^{11}$ particles for the dimension calculations, and
(including probes) $4\times 10^{11}$ particles for the average profile.

Both periodic and reflective boundary conditions have been implemented on the
sides of the channel (the periodic boundary condition is sometimes referred to
as ``cylindrical'').  The reflective boundary conditions is achieved as above:
the cluster is grown in a channel of double width and
periodic boundary conditions, and for each deposited particle we deposit also
its mirror image.  At the end one of the images was discarded.  By the
conventions used in this paper the channel is given by the range $-w/2<y<w/2$
and the clusters grow (macroscopically) in the positive $x$ direction.


\emph{Fractal dimension of channel DLA.}
The fractal dimension is measured through the density.  The average density
scales with the width $w$ of the channel, with exponent given by the
co-dimension:
\be
\rho(w) \sim w^{D-2}
\ee
To avoid transients, we discarded the first and last $3w$ long section of the
clusters, and measured the density (number of particle centers per area) on
the remaining middle section.

We generated clusters of $8\times 10^6$ to $32\times 10^6$ particles in
channels of width $w=50,100,200,500,1000,2000,5000$ particle diameters.  For
each width the number of clusters grown ranged from a few hundred to a few
thousand, with more and larger clusters necessary for large widths, to achieve
comparable statistical confidence in the average density.

\begin{figure}
\resizebox{\figurewidth}{!}{\includegraphics*{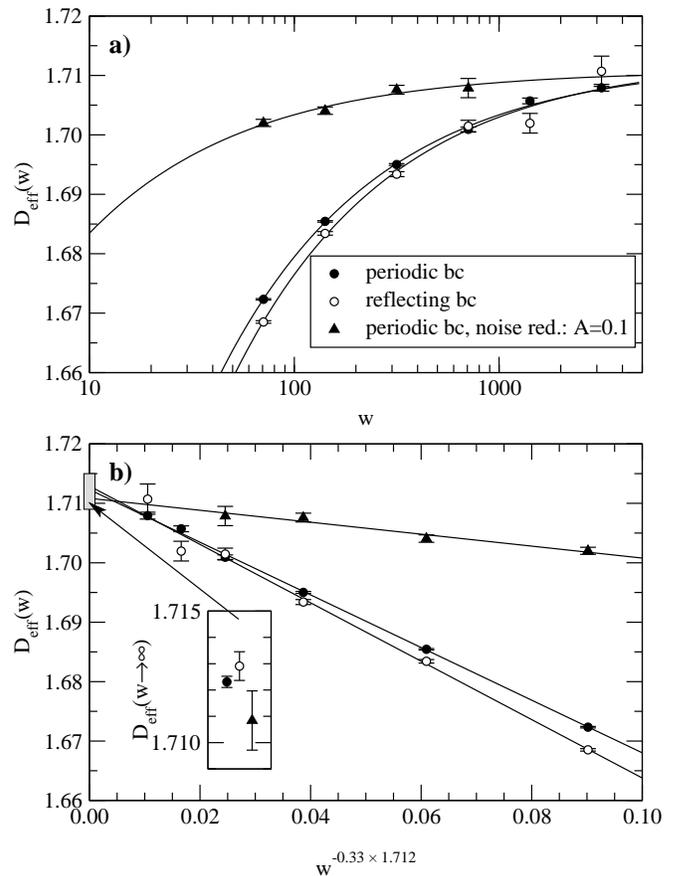}}
\caption{\label{fig:dim}
\textbf{a)} The (effective) fractal dimension as a function of channel width
$w$.  Circles correspond to the original model with periodic or reflecting
boundary conditions, the triangles are made with off-lattice noise reduction
\cite{ball02pre}. For the three cases the ensemble size (and therefore the
statistical uncertainty) was different. The curves correspond to the fitted
lines in the b) panel. \textbf{b)} The finite size scaling plot of the
same quantities, see Ref.~\cite{somfai99}. \textbf{Inset:} All data is
consistent in the $w\to\infty$ extrapolation with the dimension $D = 1.712\pm
0.002$. (In the inset the data is shifted horizontally for clarity.)
}
\end{figure}

Figure~\ref{fig:dim} shows the width ($w$) dependence of the effective fractal
dimension $D_\text{eff} = 2+d\ln\rho / d\ln w$.  The fractal dimension tends
to $D=1.712\pm 0.002$, independent of the choice of boundary conditions.
Off-lattice noise reduction \cite{ball02pre} does not change the dimension,
but accelerates the convergence to its asymptotic value.

\emph{Average profile.}
Analytical solutions for unbranched Laplacian growth in a channel with
reflective boundary conditions have been known for a long time.  One can find
solutions which translate a fixed profile along the channel in time.  In the
absence of surface tension, these solutions form a one-parameter family,
called Saffman-Taylor (ST) fingers \cite{saffman58}. They are parametrized by
the asymptotic ratio $\lambda$ of the widths of the finger ($w_\text{finger}$)
and the channel ($w$), and have the profile:
\be
x(y) = \frac{w(1-\lambda)}{2\pi}\log\left[\frac{1}{2}\left(1 + \cos \frac{2\pi
y}{\lambda w}\right)\right]
\ee
Of these solutions $\lambda=1/2$ is the most important, because in related
experiments \cite{saffman58} this profile has been observed in the limit of
vanishing surface tension. Analytical calculations
\cite{combescot86,shraiman86,hong86} show that surface tension---a singular
perturbation---selects a discrete set of finger solutions (only one of which
is linearly stable), which all converge to the $\lambda=1/2$ ST-finger in the
limit of zero surface tension.


It has been suggested \cite{arneodo89}  that the $\lambda=1/2$ ST-finger
solution also models the average profile both of the unstable (highly
branched) Hele--Shaw fingering and of the DLA growth in a channel with the
corresponding reflective boundary conditions.  The profile was defined as a
level set of the ensemble averaged mass density, and for the experimental
Hele--Shaw profiles half the maximum level was used.  For DLA growth it was
later shown \cite{arneodo96} that the level set at 0.5 maximum matches the
width of $\lambda=0.56$,  whilst the best match to the $\lambda=1/2$ profile
came from the level set at 0.6 maximum.  Outside that range the authors of
Ref.~\cite{arneodo96} concluded they could match level sets only to finger
widths but not to the full shape of any ST-finger.

Here we use a different kind of finger averaging, which does not have any
fitting parameter (eg. height of level set), as follows. 
In the conformal map method, we directly average the map. That is, we choose a 
set of points on the unit circle in the mathematical plane, and repeatedly map
to the physical plane. The position of the image points averaged over the
different maps, that is, over the different clusters that we have generated,
is a reasonable alternative to the ensemble average of
\cite{arneodo89,arneodo96}.

\begin{figure}
\resizebox{\figurewidth}{!}{\includegraphics*{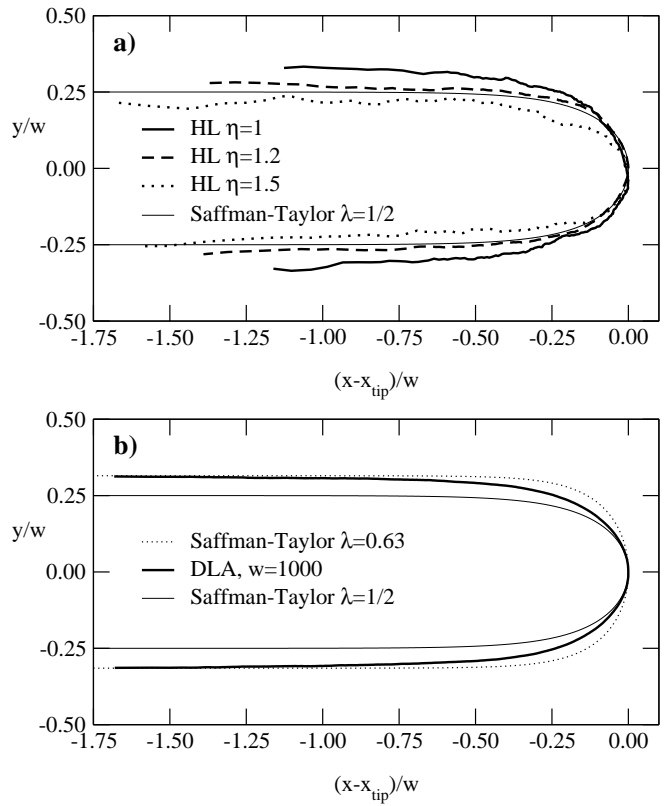}}
\caption{\label{fig:finger}
\textbf{a)} The average conformal map, generated with iterated conformal maps
for $\eta=1$, $1.2$ and $1.5$. The profile for $\eta=1.2$ comes the closest to
the ST-finger solution for $\lambda=1/2$.
\textbf{b)} The average map of DLA clusters grown with random walking
particles in a 1000 particle-diameter wide reflective channel. The profile
does not follow any ST-finger solution.
}
\end{figure}

In Fig. \ref{fig:finger}(a) we show the average conformal map generated this
way.  We see that the average map for DLA, $\eta=1$, does not correspond to
the ST result for $\lambda=1/2$, but we get a good match to it for the DBM
growth at $\eta=1.2$.
This is an interesting result, especially in the context of recently proposed
equivalences between DBM models with generalized local spatial cutoff. In that
framework \cite{ball02prl,ball03pre} a higly ramified viscous finger with
simple surface tension cutoff corresponds to standard (fixed size cutoff) DBM
with $\eta\approx 1.2$. Here we observe that the
\emph{non-branching} ST-finger solution is very similar to the \emph{conformal
average} of DBM clusters of the same $\eta$.

To use DLA grown with random walking particles in a channel, 
we  need only construct the conformal map from the complex unit
circle to the perimeter of each cluster, and  take the average of these
maps, as above. 
The conformal map is obtained
numerically by the following method \cite{somfai99}.  We send $M$ probe
particles to the frozen cluster, record their impact position, and discard
them.  These points correspond to $M$ uniformly distributed points on the unit
circle.  The landing positions of the probe particles are labeled
topologically as one encounters them when tracking the perimeter of the
cluster.  Finally the $m$-th point is assigned to the angle $2\pi m/M$ of the
unit circle.  This angle has an error of the order of $M^{-1/2}$, which
vanishes for large $M$.

We measured the average conformal map on channels with reflective boundary
conditions, widths ranging from 10 to 2000 particle diameters. For each width,
we grew $10^5$ short clusters (only about $10w$ long), and probed each with
$10^5$ test particles. In addition, for a few selected widths we probed $10^4$
clusters with $10^6$ probes each.

The average map for a wide channel generated this way
is shown in Fig.~\ref{fig:finger}(b). 
The curvature of the tip is consistent with that of the $\lambda=1/2$
ST-finger (the measured curvature of the DLA profile for $w=1000$ or $2000$
corresponds to $\lambda=0.51\pm0.03$).  The asymptotic width, however, is
larger.  The average map significantly differs also from the ST-finger of
matching asymptotic width.

\begin{figure}
\resizebox{\figurewidth}{!}{\includegraphics*{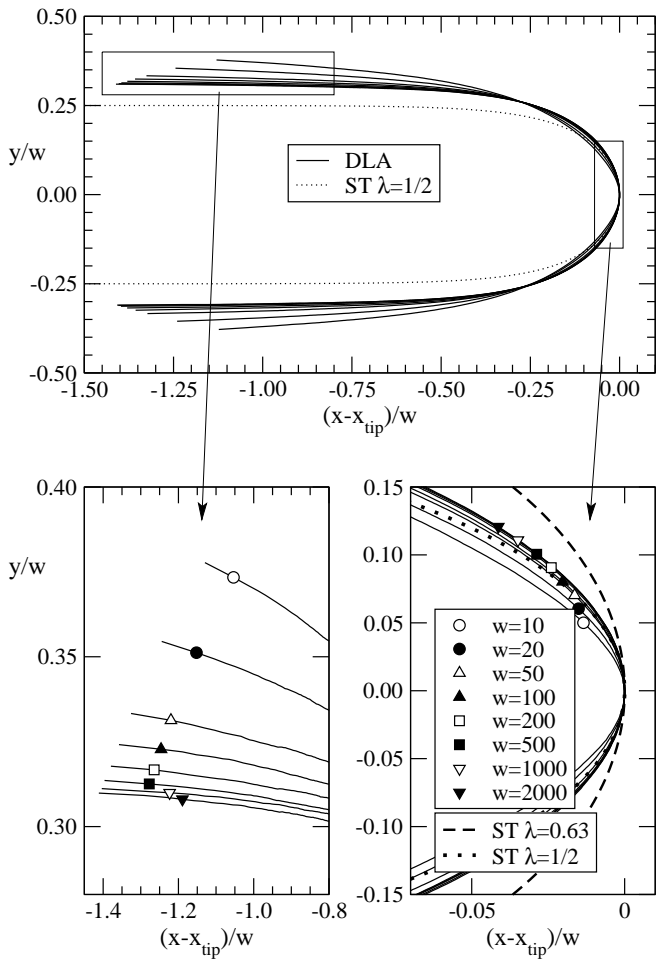}}
\caption{\label{fig:finger-w}
  The $w$ dependence of the average conformal map, rescaled onto a unit wide
channel. The tip and the tail are magnified on the bottom panels.
}
\end{figure}

The average conformal map, rescaled onto a unit wide channel, shows strong
dependence on the channel width. This is shown on Fig.~\ref{fig:finger-w} as a
function of reduced $x$ position relative to the tip,
$\xi=(x-x_\text{tip})/w$.  Details of the tip and tail regions show that these
are clearly not consistent with a common asymptotic ST-finger shape.

\begin{figure}
\resizebox{\figurewidth}{!}{\includegraphics*{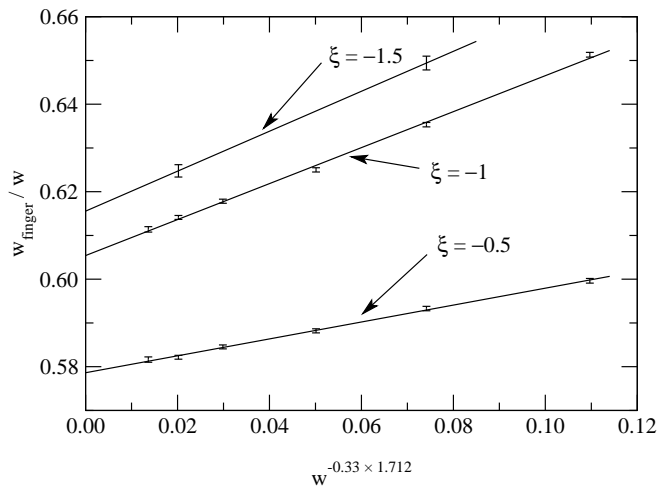}}
\caption{\label{fig:finger-scale}
  Finite size scaling of the finger filling ratio $w_\text{finger}(\xi)/w$,
measured at a few fixed $\xi$ values. Only channels $50\le w \le 2000$ were
included in the fits.  The finite size scaling gives the extrapolation
$w\to\infty$. For the second extrapolation, $\xi\to -\infty$, we can only state
that $w_\text{finger}/w \gtrsim 0.62$.
}
\end{figure}

We performed a finite size scaling on the $w$ dependence of the finger width.
The filling ratio of the finger $w_\text{finger}/w= [y_+(x)-y_-(x)]/w$ was
measured at selected $\xi$ values:  $\xi= -0.5$, $-1$, and $-1.5$, and is
plotted on Fig.~\ref{fig:finger-scale}. The finite size scaling exponent was
found to be $\nu=0.33$, same as for other quantities
\cite{somfai99,ball02pre}.  The most interesting is the second extrapolation:
$\xi\to-\infty$.  We only have 3 points for this, so it is only reasonable to
give a lower bound: $w_\text{finger}/w \gtrsim 0.62$.


In summary, using large scale random walker based simulations we have shown
that the fractal dimension of DLA in a channel---with either periodic or
reflective boundary conditions---is the same as in radial geometry. This is a
great simplification compared to earlier claims of boundary condition
(geometry) dependent fractal dimension. Second, using both iterated conformal
maps and random walker based simulations, we measured the average profile of
the clusters, defined by the average conformal map, and compared them to
ST-finger solutions of the corresponding continuum problem. The averaged DLA
profile is reminiscent but distinct from the ST-fingers, while the average
profile of DBM clusters with $\eta=1.2$ are rather similar to the ST-finger
with $\lambda=1/2$.

\begin{acknowledgments}
We are indebted to Dave Kessler for the suggestion of measuring the average
conformal map.
This research has been supported by the EC under Contract No.
HPMF-CT-2000-00800.
The computing facilities were provided by the Centre for Scientific Computing
of the University of Warwick, with support from the JREI.
\end{acknowledgments}

\bibliography{dlaref}

\end{document}